\documentclass[runningheads,a4paper]{llncs}

\usepackage{amssymb}
\usepackage{amsmath}
\usepackage{graphicx}

\newcommand{\keywords}[1]{\par\addvspace\baselineskip
\noindent\keywordname\enspace\ignorespaces#1}

\begin{document}

\mainmatter  

\title{Putting the SC in SCORE: Solar Car Optimized Route Estimation and Smart Cities}

\titlerunning{Putting the SC in SCORE}

\author{Mehrija Hasicic\inst{1} \and
Harun Siljak\inst{2}}
\institute{Electrical and Electronics Engineering Department, International Burch University Sarajevo, Bosnia and Herzegovina
\and CONNECT Centre, Trinity College, The University of Dublin, Ireland\\
\email{mehrija.hasicic@ibu.edu.ba},\,\email{harun.siljak@tcd.ie}}

\toctitle{Putting the SC in SCORE: Solar Car Optimized Route Estimation and Smart Cities}
\tocauthor{-}
\maketitle

\begin{abstract}
Solar exposure of streets and parking spaces in dense urban areas varies significantly due to the infrastructure: buildings, parks, tunnels, multistorey car parks. This variability leaves space for both real-time and offline route and parking optimization for solar-powered vehicles. In this chapter we present Solar Car Optimized Route Estimation (SCORE), our optimization system based on historic and current solar radiance measurements. In addition to the comprehensive review of SCORE, we offer a new perspective on it by embedding it in the bigger picture of smart cities (SC): we analyze SCORE's relationship with the smart power generation and distribution systems (smart grid), novel transportation paradigms and communication advancements. While the previous work on SCORE was focused on technical challenges which are described in the first part of this chapter (optimization, communication, sensor data collection and fusion), here we proceed with a systemic approach and observe a SCORE-equipped unit in the near-future society, examine the sustainability of the model and possible business models based on it. We consider the problem of vehicle routing and congestion avoidance using incentives for users on non-critical journeys and co-existence of SCORE and non-SCORE using vehicles. Realistic pointers for SCORE-aware design of infrastructure are also given, both for improved data collection and improved solar exposure while considering trade-offs for non-SCORE users.

\end{abstract}

\section{Introduction}

When we introduced Solar Car Optimized Route Estimation (SCORE), we were solving the problem of a single, possibly unique solar vehicle optimising its route in a \emph{conventional} city. In this chapter we examine the role of such a smart system, devised to find a sunny route (and a sunny parking lot) for a solar car, in a larger smart eco-system, one of a \emph{smart} city.

We recognise that solar vehicles of different sizes and different types of hybridization \cite{mulhall_solar-assisted_2010,arsie_optimal_2006} have been proposed, as well as options to make existing cars hybrid solar vehicles \cite{luca_aftermarket_2018}, and that solar cars are a part of a larger scheme in the vision of smart city, smart transport, and any future involving fighting climate change \cite{popiolek_multi-criteria_2016,joshi_sustainable_2018}. How to fit SCORE-guided solar vehicles in a smart transportation system hosting other electric and hybrid vehicles, how to fit it in a smart grid system which is sensitive to electric cars being plugged-in, and how do they fit in the landscape of smart city's parking lots which should serve as charging stations? These are the questions we share our opinions on, after presenting the idea of SCORE in brief.

\section{SCORE}
Solar Car Optimized Route Estimation (SCORE) system is a system that deals with the data acquisition, software and hardware developed propositions for route and parking selections for a (hybrid) solar vehicle. Detailed explanation of the system is given in the following subsections.

\subsection{System Description}
\begin{figure}
    \centering
    \includegraphics[width=0.8\textwidth]{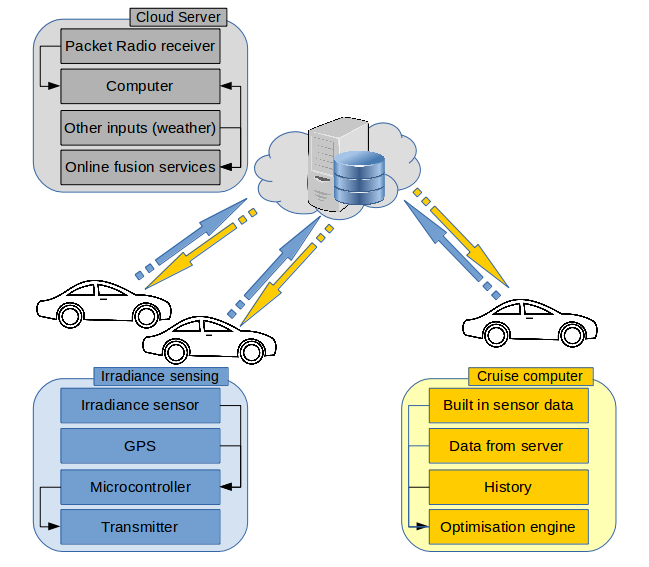}
    \caption{Overall structure of SCORE \cite{hasicic_criteria_2017}}
    \label{Fig.1:}
\end{figure}

The SCORE is built up from three separate parts shown in Fig. \ref{Fig.1:} and listed below:
\begin{itemize}
    \item irradiance sensing - mobile sensor that transmits solar radiance from the streets in real time. These transmitter should, preferably be placed on the frequently moving cars or could be placed on fixed locations in  the streets;
    \item cloud servers - server for data fusion that collects the data from the field and the third party and combining it with offline data and sending it to the device embedded in the cars in form of matrix;
    \item cruise computer - embedded computer unit in the solar car that combines received data with its own readings and normalizes the data using built in light sensor. Proposed route changes dynamically based on weather updates.
\end{itemize}

Prototype implemented in this research uses only historical data from the cloud due to the limitations of self-built solar car, and due to the fact that only one car in this case uses SCORE. 

\subsection{Data Collection and Management}
As previously mentioned SCORE is designed to work with both, online and offline data. Two ways of collecting it are proposed: weather forecast and mobile sensor installed on solar and regular cars (i.e. taxi, delivery tracks etc.).

\subsubsection{A priori data}

\begin{figure}
    \centering
    \includegraphics[width=0.8\textwidth]{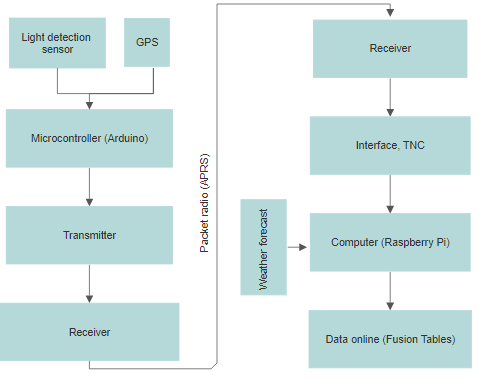}
    \caption{(a) Mobile device structure collects the data from GPS and light detection sensor and transmits it to the (b) Central server structure which process the data and stores it online, available for end user (EV drivers) to download it
    \cite{hasicic_criteria_2017}}
    \label{Fig.2:}
\end{figure}

Architecture of the device (using components compact enough to be placed on any car to collect the data without any needs for route or vehicle customization) is shown in Fig. \ref{Fig.2:}(a). If placed on taxis or delivery tracks this device could collect huge amounts of data random in time, date and location, opening the space for big data analytics as well.

Proposed way in this research for sending the data is APRS (Automatic Packet Reporting System) as the easiest packet radio implementation. However, any wireless protocol can be used for the same purpose as well as proprietary and private frequencies.

The server, shown in Fig. \ref{Fig.2:}(b), receives the data and converts it to text using common sound card and appropriate software. This data and data from CAD (Computer Aided Design) and GIS (Geographic Information System) is put in tabular form online. CAD data is obtained by simulating sun movement in 3D model of the street, while GIS data is provided by GIS services measuring solar radiance of different areas.

\begin{figure}
    \centering
    \includegraphics[width=0.8\textwidth]{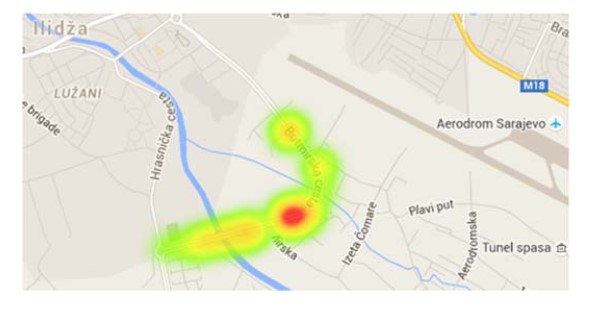}
    \caption{Visualization of the data from Fusion Table for local map showing irradiation on the route from our University to the city center. Irradiation has been showed by the colours, red representing iraddiation 1 - very sunny route and green representing low iraddiation - route being in the shadows of near buildings or trees. \cite{hasicic_sensor_2016}}
    \label{Fig.3:}
\end{figure}

Shown in Fig. \ref{Fig.2:} is an exemplary scheme of the technologies proposed in this research which can always be replaced by some existing alternatives. Fusion tables are proposed due to the ability of presenting the data visually as shown in Fig. \ref{Fig.3:}.

\subsubsection{Real-time data}
\begin{figure}
    \centering
    \includegraphics[width=0.8\textwidth]{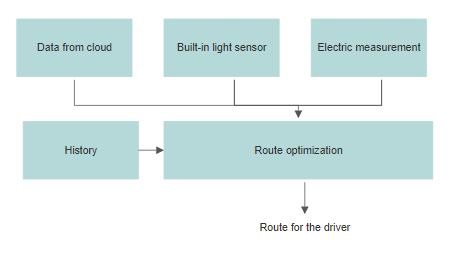}
    \caption{Embedded car computer structure \cite{hasicic_sensor_2016}}
    \label{Fig.4:}
\end{figure}

I/O architecture of the embedded computer placed in the solar car is shown in Fig.\ref{Fig.4:}. To optimize the route system uses routes from user's history, sensor fusion data from the cloud where the most resent data is the one that matters the most , measurements from the solar panels and battery and measurement from the built-in light sensor. 

Measurements from the built-in sensor are used as corrective measurements of the real output and predicted one, where error rate is used to calibrate all others predicted values. On the other hand, electrical measurements are used to predict energy consumption and cost of every available route from beginning to end destination. All the facts and measurements taken in consideration result in proposition of the best route and parking with high solar irradiation. Combination of the multiple objectives (route and parking) deserves more attention in future research, as well as variation in parameters of the optimisation problem.

\subsection{Theoretical considerations}

Theoretical considerations must be taken into account when considering two things: solar irradiation data and optimization. Since we have proposed two types of data collection each has its own considerations to be analyzed:
Online data –  normalized data represented from 0 (no radiance) to 1 (maximum radiance) with a time-stamp denoting the time expressed in hours with beginning of the year as the reference point;
Offline data – data predicted for particular area using weather forecast, CAD and GIS.
Final irradiation value, $r$, is calculated using normalized irradiation values from previously mentioned online and offline data, $r_{on}$ and $r_{off}$, expressed as:
\begin{equation}
    r=r_{on}\cdot a + r_{off}\cdot(1-a)
\end{equation}
where $a$ is expressed as:
\begin{equation}
    a=\exp\left(\frac{=(t_{curr}-t_{meas})^2}{100000}\right)
\end{equation}
and $t_{curr}$ and $t_{meas}$ are time expressed in hours from the beginning of the year, where $t_{curr}$ stands for current time and $t_{meas}$ represents the time of last measurement of the data. Denominator is chosen empirically.

Dijkstra's optimization algorithm is chosen for the route optimization where positive weights of edges and nodes needed to be decided. All main crossroads in Sarajevo were chosen to be nodes. Beside the length of the road which is obvious factor, we also needed to take in consideration converted solar energy while on the road. Car specification that affect the converted solar energy are: \begin{itemize}
    \item motor power - 11 kW
    \item panel area - $2\times 0.726 \text{m}^2$
    \item panel efficiency - 18 percent
    \item received power per square meter of the panel - 957 W/$\text{m}^2$ under maximum radiance, which results in 30 percent of radiance is reflected.
\end{itemize} In this research for the purpose of developing prototype we were doing calculations and server serves only one car at the time. the idea is to develop it further more to be functional for unlimited number of solar cars at the same time. The figures in the specification are relevant as they define the efficiency of solar energy conversion and the requirements for the operation of the vehicle. With greater conversion efficiency expected in new car development, SCORE's effect will be augmented as well.

Solar irradiation of the road segments is taken as the arithmetic mean of values taken from beginning and end node. This crude approximation is used for the prototype development but will be avoided in case when mobile sensors are installed on cars and start cruising the city.

As for the parking lot, both irradiation of the parking and its distance from the destination are taken in consideration. Parking with the highest ratio of irradiation and distance is the one that is going to be proposed to the user. Whether irradiation or the distance is more important can be decided by user by taking either one or another to a non-unit power.

All previously mentioned calculations are done on server and client only gets the proposed route and parking. However maps are updated to reflect changes in reality and update of calculated routes are regularly conducted since weather may vary significantly, especially on longer trips. Frequency of updates can be variable (e.g. urban trips vs. intercity trips).

\subsection{Implementation}
\subsubsection{Implementation of sensor data collection and the server}
As previously mentioned, mobile device developed in the scope of this research is compact and autonomous which allows its placement on the car without customization of the car itself. For the data collection purpose it can be placed on any vehicle, not only on the solar cars.

APRS is proposed as transmission channel for delivering GPS position and sensor data to the terminal node. It has been used in monitoring systems for the same purpose \cite{bello2007design}. For the transmission we have used amateur radio bands, however different frequencies and wireless protocols can be used for the different implementation of the SCORE.

Since we are interested not only in historic data but also in the real time data we need to be able to constantly update and communicate with the server.

We did not need large computing power for the server so we implemented it on Raspberry Pi 2, even so the same could be done on cloud as well. Enabling all parties to access fusion tables was the main reason to store them on the cloud, this allows client to access them on their smart phones and/or computers as well.

As the SCORE system scales up and more vehicles adopt it, various approaches for congestion avoidance can be employed: caching on a cloud-edge scheme, local broadcasting, etc.

\subsubsection{Implementation of the optimization client}
Beside previously mentioned prototype done on microcontroller we have also implemented testbed for algorithm testing in MATLAB (simulation of the server-client communication). Graphical user interface (GUI) in MATLAB is shown in Fig. 5.
\begin{figure}
    \centering
    \includegraphics[width=0.8\textwidth]{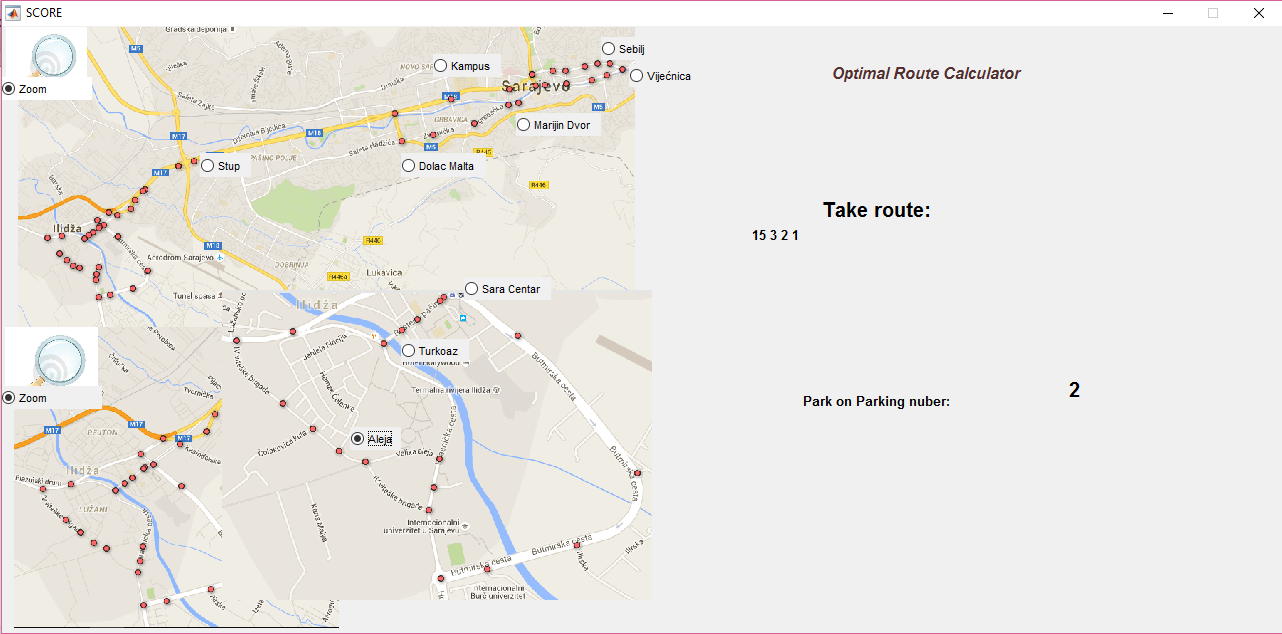}
    \caption{GUI in MATLAB. Each red dot is a node that has been assigned a number. Using those numbers computer shows us the optimum route. Data with parking slots near every node is provided to the software, based on equations it calculates and shows you nearest parking with highest irradiation too. }
    \label{Fig. 5:}
\end{figure}

As it was previously mentioned, in the beginning stage of our research we have planned to compute car computer on ARDUINO. However due to the significantly more memory, needed to keep whole matrix representing the graph in working memory of the processor, the car computer prototype was built on TI's ARM Cortex-M4F based TM4C123G LaunchPad.
\begin{figure}
    \centering
    \includegraphics[width=0.8\textwidth]{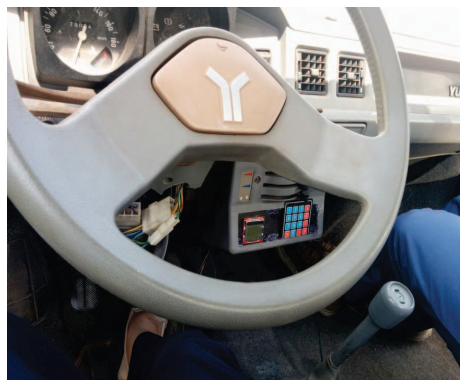}
    \caption{The navigation client placed in the car \cite{bilic_practical_2016} }
    \label{Fig. 6:}
\end{figure}

The car computer has the keyboard, used for entering the destination through choosing the node number, and the display used for showing the route through the node numbers (crossroads) client needs to pass in the exact order as showed. Simple implementation we did, shown in Fig. 6, gets the matrix data through wireless or USB debug cable and optimizes using Dijkstra's algorithm and displays the set of nodes connecting starting and end node.

\section{Putting the smart city (SC) into SCORE}

Solar energy resources are, alongside other renewables, instrumental element of smart cities \cite{wang2016efficient}. However, the typical solar power discourse in smart cities is one of homes and power stations, fixed objects with a fixed photovoltaic setup. What changes when we include solar cars into the equation?

In the case of a single solar vehicle using SCORE, not much changes in the smart city: one driver will use routes and parking lots recommended by SCORE, while the only effect that can be felt in community is one coming from the size of the mobile sensor - community updating solar irradiance data online (which, as we have already noted, would be independent from solar cars themselves, and would be placed on vehicles performing public services).

It is much more interesting to examine the situation when a growing community of drivers uses SCORE: a significant portion of population uses hybrid solar cars, alongside those who use the non-solar hybrid cars, and maybe a portion of fossil fuel cars as well in the transitional period. In this scenario, the question of other vehicles occupying best routes for solar cars may be a hindering issue (and a frustrating one for solar drivers). This is where SCORE needs integration with other software and hardware platforms in the future vehicular networks: be it in our contemporary navigation, mapping and route recommendation systems, or in the autonomous driving systems of the future, SCORE input would give recommendations to non-SCORE drivers to use other routes when SCORE drivers are expected to use their well-insolated routes. Direct incentive is jam avoidance, but a different in-traffic economy could be built as well, with incentives like good parking spots or free charging station time for cars that choose to go with less attractive routes, to balance the traffic. 

This is what the citizens of the smart city (at least those who drive) see as the effect of introduction of SCORE. What is the effect noticed by sensors in the power grid and the dispatchers in the smart grid of the smart city? SCORE-enabled vehicles, being at least partially solar, would operate as gridable elements, contributing power into the system or taking it out of the system while plugged in, depending on the battery state. The effect of gridable vehicles is studied in literature \cite{su_survey_2012,saber_efficient_2010}, but the vehicles generally observed are those who charge solely from the grid, i.e. not having own renewable source as solar vehicles do. An additional renewable source connected to the grid can offer a new degree of freedom for the smart grid. The idea behind smart parking for electric vehicles revolves around smart power management of cars connected to grid\cite{honarmand_optimal_2014}. If the solar component is included, it is the parking lot itself with solar panels \cite{nunes_use_2016}, but here we bring the solar panels to places that need them (and places that can use them well based on their insolation), encouraging the drivers to park there.

In both of these scenarios, it is the relationship between a SCORE-user and a non-SCORE entity that determines the success of implementation. 

A SCORE-user wants a (1) satisfying travel and (2) convenient parking. The adjectives used here are vague: a satisfying travel could be the one that's as short as possible (if that is the priority) or the one that is as cheap as possible (if that is the priority), or a mix of the two. Similarly, the parking could be the most economical one (charging the most in high insolation conditions) or simple the most comfortable (i.e. closest to the destination). From a game-theoretic perspective, the question is how conflicting these interests of the SCORE-user are with those of other, non-SCORE users.

An electric vehicle (EV) driver could state the same vague requests for travel and parking as our SCORE-user, but the mechanics behind them would be different, as the EV driver's power budget can never increase "for free". This simplifies the optimization for the EV driver, as they aim to minimize the losses, not able to expect a power contribution from the outside. This in practical conditions means that often the optimal routes for SCORE and non-SCORE drivers will not be the same, and they do not have to directly compete for them. In cases where they do compete, the solution can be brought by either congestion pricing \cite{washburn2009helping} or elaborate pricing schemes that have been developed for electric vehicles \cite{alizadeh2016optimal}.

The other non-SCORE entity in the game is the grid, and it plays the same game with the EV driver as it does with a SCORE driver, with the difference that a SCORE vehicle would serve as a mobile renewable source if parked for a long time period, a benefit not seen from the regular EV (which at best would serve as a mobile battery). If we extend this player's scope to include the general infrastructure, it can include the parking lot planning--and insightful planning of parking lots with natural sunlight would come as a new factor, opposing the previous architectural desire to create as much shade as possible in parking spaces \cite{mcpherson2001sacramento}. 

\section{Conclusions}

SCORE is a convenient framework for a small group of solar vehicles finding their way in a modern town. Only with the smart, connected cities of the future will it see its full potential integrating into smart transport and smart grid of the city alike. A social benefit to the ecosystem and providing the service to its user are key benefits of implementing the SCORE into SC. We argue that it will complement other smart transport solutions in a natural fashion, without disruptions and deadlocks in optimization.

SCORE's promise of integration lies in its dual flexibility and robustness: it is responsive as it tracks all relevant changes in the environment, and with its multi-criterial extensions (parking, route, etc) it can provide a wide spectrum of solutions catering for different needs (criticality, extreme pollution decrease, daily congestion management).

\section{Acknowledgements}

We thank Mr Damir Bilic (Mälardalen University, Sweden) for his invaluable contribution in developing SCORE, and useful discussions in preparation of this chapter.

\end{document}